\documentclass{article}
    \PassOptionsToPackage{numbers, compress}{natbib}
    \usepackage[final]{neurips_2024}

\usepackage[utf8]{inputenc} % allow utf-8 input
\usepackage[T1]{fontenc}    % use 8-bit T1 fonts
\usepackage{hyperref}       % hyperlinks
\usepackage{url}            % simple URL typesetting
\usepackage{booktabs}       % professional-quality tables
\usepackage{amsfonts}       % blackboard math symbols
\usepackage{nicefrac}       % compact symbols for 1/2, etc.
\usepackage{microtype}      % microtypography
\usepackage{xcolor}         % colors
\usepackage{kotex}          % korean
\usepackage{hyperref}
\usepackage{amsmath}
\usepackage{adjustbox}
\usepackage{multirow}
\usepackage{color}
\usepackage{graphicx}
\usepackage{caption}
\usepackage{subcaption}  
\usepackage[none]{hyphenat}
\usepackage{colortbl}
\usepackage{makecell}

\title{A Gradient Accumulation Method for Dense Retriever under Memory Constraint}

\author{%
  Jaehee Kim$^{1}$ \quad Yukyung Lee$^{2}$ \quad Pilsung Kang$^{1}$\thanks{indicates corresponding author} \\
  $^{1}$Seoul National University \quad $^{2}$Boston University\\
  \texttt{\{jaehee\_kim, pilsung\_kang\}@snu.ac.kr} \\
  \texttt{ylee5@bu.edu}
  }

\newcommand{\modelname}{\textsc{ContAccum}}
\newcommand{\longmodelname}{Contrastive Accumulation}
\newcommand{\ga}{GradAccum}
\newcommand{\gc}{GradCache}
\newcommand{\gradimbalproblem}{\textit{gradient norm imbalance problem}}
\newcommand{\gradnormratio}{\text{GradNormRatio}}

\begin{document}

\maketitle

\begin{abstract}
  % The abstract paragraph should be indented \nicefrac{1}{2}~inch (3~picas) on
  % both the left- and right-hand margins. Use 10~point type, with a vertical
  % spacing (leading) of 11~points.  The word \textbf{Abstract} must be centered,
  % bold, and in point size 12. Two line spaces precede the abstract. The abstract
  % must be limited to one paragraph.
    InfoNCE loss is commonly used to train dense retriever in information retrieval tasks. It is well known that a large batch is essential to stable and effective training with InfoNCE loss, which requires significant hardware resources. Due to the dependency of large batch, dense retriever has bottleneck of application and research. Recently, memory reduction methods have been broadly adopted to resolve the hardware bottleneck by decomposing forward and backward or using a memory bank. However, current methods still suffer from slow and unstable training. To address these issues, we propose \longmodelname\ (\modelname), a stable and efficient memory reduction method for dense retriever trains that uses a dual memory bank structure to leverage previously generated query and passage representations. Experiments on widely used five information retrieval datasets indicate that \modelname\ can surpass not only existing memory reduction methods but also high-resource scenario. Moreover, theoretical analysis and experimental results confirm that \modelname\ provides more stable dual-encoder training than current memory bank utilization methods.
\end{abstract}

\section{Introduction}
\label{sec:introduction}
Dense retriever aims to retrieve relevant passages from a database in response to user queries with neural networks \cite{retrieval-survey}. \citet{dpr} and \citet{orqa} introduced the in-batch negative sampling for training dense retriever with InfoNCE loss \cite{code}, where relevant passages from other queries in the same batch are utilized as negative passages. This negative sampling strategy has been widely adopted in subsequent dense retriever studies, including supervised retriever \cite{dpr, pair, rocketqa, ance}, retriever pre-training \cite{condenser, cocondenser, seed, contriever, promptagator}, phrase retriever \cite{densephrases, npm}, and generative retriever \cite{genret, lm-indexer}. Training dense retriever with InfoNCE loss drives the representations of queries and relevant passages closer and pushes the representations of unrelated passages apart, which can be seen as a form of metric learning \cite{metric-learning}.

Many dense retriever methodologies utilize large batch to incorporate more negative samples \cite{ance, condenser, rocketqa, contriever}. Theoretically, it has been demonstrated that more negative samples in InfoNCE loss lead to a tighter lower bound on mutual information between query and passage \cite{code}. Empirical studies have shown that the dense retriever performs better with large batch \cite{rocketqa, retrieval-survey, trisampler}. However, training with large batches requires high-resource, posing a challenge for dense retriever research and applications.

A line of research has focused on overcoming these limitations by approximating the effects of large batch sizes. Gradient Accumulation (\ga), a common method for approximating large batch, reduces memory usage by splitting the large batch into smaller batches. However, \ga\ has limitations in the context of InfoNCE loss because it reduces negative samples per query by the smaller batch \cite{gradcache}. To overcome the limitation of \ga, \citet{gradcache} proposed the Gradient Cache (\gc), which approximates large batch by decomposing the backpropagation process and adapts additional forwarding process for calculating gradients. However, \gc\ has limitations, including significant additional training time due to computational overhead and the inability to surpass high-resource scenario where accelerators are sufficient to train large batch. Additionally, pre-batch negatives \cite{densephrases} caches passage representations from previous steps to secure additional negative samples, but it also shows unstable train and marginal performance gain.

In this study, we propose \textbf{\longmodelname\ (\modelname)}, which demonstrates high performance and stable training under memory constraints. \modelname\ leverages previously generated query and passage representations through a memory bank, enabling the use of more negative samples. Our analysis of the gradients reveals that utilizing a memory bank for both query and passage leads to stable training. The specific contributions of this study are as follows: 
\begin{itemize}
    \item We propose \modelname, a method utilizing a dual memory bank strategy that can outperform not only existing memory reduction methods but also high-resource scenario in low-resource setting.
    \item We show that our method is time efficient, reducing the training time compared to existing memory reduction methods. 
    \item We demonstrate the cause of training instability in existing memory bank utilization methods through mathematical analysis and experiments, showing that the dual memory bank strategy stabilizes training.
\end{itemize}

\section{Related works}
\label{sec:related works}
\begin{figure}[h]
    \centering
    \includegraphics[width=0.7\columnwidth]{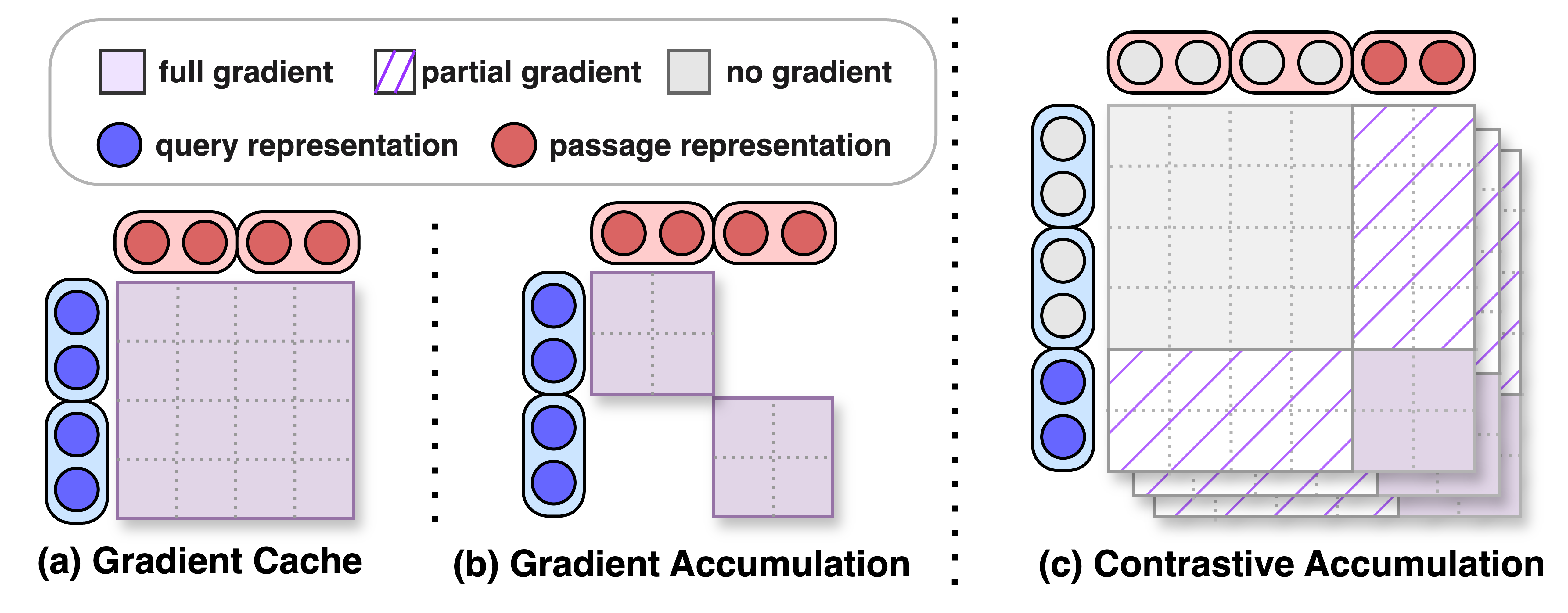}
    \caption{\textbf{Illustrations of \modelname\ and Comparative Methods.} The illustrations show a total batch size (\(N_\text{total}\)) of 4, a local batch size (\(N_\text{local}\)) of 2, and a memory bank size (\(N_\text{memory}\)) of 4. (a) \gc\ uses \(N_\text{total} - 1\) negative passages. (b) \ga\ uses \(N_\text{local} - 1\) negative passages. (c) \modelname\ leverages \(N_\text{local} + N_\text{memory} - 1\) negative samples, more than \(N_\text{total} - 1\).}
    \label{fig:comparison-contaccum}
\end{figure}
\subsection{Memory reduction in information retrieval} %yukyung
\ga\ is the most common method to address memory reduction problem. By using \ga, gradients of the total batch can be stored by sequentially processing local batches through forward and backward passes, even when the total batch cannot be processed at once. However, as shown in Figure \ref{fig:comparison-contaccum} (b), \ga\ is not a proper memory reduction method for the in-batch negatives, as it uses fewer negative samples than the total batch. We will discuss the limitation of \ga\ for contrastive learning in detail in subsection \ref{subsect:preliminary}.

\gc\ reduces memory usage in contrastive learning by decomposing the backpropagation process. Specifically, as shown in Figure \ref{fig:comparison-contaccum} (a), it calculates the loss without storing activations during the forward pass using the total batch. Then, it computes and stores the gradient from the loss to the representations. Next, it performs additional forward passes for the local batch to store activations and sequentially calculates gradients from each representation to the model weights. This allows \gc\ to use the same number of negative samples as the total batch, approximating the performance of the total batch. However, \gc\ cannot surpass the performance of high-resource scenario because it uses the same number of negative samples. Also, \gc\ requires a significant amount of time due to the complex forward and backward processes. 

\subsection{Memory bank}
The memory bank structure for metric learning was initially proposed for the vision domain, where it stores representations generated by the encoder in previous batches \cite{memory-bank, memory-bank-neg}. Combined with the NCE loss \cite{nce}, memory bank structures have been widely used to train uni-encoder vision models \cite{mocov1, mocov2, xbm}. However, directly adapting this approach to information retrieval tasks, where a dual-encoder structure is commonly used, is challenging. This is due to several factors: %이 부분에 너가 생각하는 이유 간단히 추가
In multi-modal settings, \citet{albef, blip} have employed momentum encoders for both image and text modalities to generate cached representations. However, these approaches do not directly address the asymmetric nature of information retrieval, where the goal is to retrieve relevant passages for a given query rather than retrieving relevant queries for a given passage.

In the information retrieval task, \citet{contriever} proposed caching representations generated by a momentum encoder \cite{mocov1}, but they only consider the uni-encoder setting. \citet{densephrases} introduced pre-batch negatives that extend the number of negative samples by caching passage representations with a memory bank in a dual-encoder setting. However, pre-batch negatives was applied only in the final few epochs of the training process due to the rapid changes in encoder representations early in training, which can cause instability when using a memory bank \cite{xbm, cbns}.

In summary, existing dense retrievers depend on in-batch negative sampling, necessitating large batch sizes and costly hardware settings. While memory reduction methods have been studied to address this, they often result in slower training or unstable training. Therefore, we propose \modelname, a memory reduction method designed to ensure fast and stable training of dense retrievers.

\section{Proposed Method}
\subsection{Preliminary: InfoNCE loss with \ga} 
\label{subsect:preliminary} 
Before introducing our method, we first examine \ga\ with InfoNCE loss. \citet{dpr} proposed training method for dense retriever using InfoNCE loss. With a batch size \(N\), dense retrievers are trained by minimizing the negative log-likelihood over all query representations (\(\mathbf{Q}\)) and passage representations. Specifically, they utilized in-batch negative sampling  (\(\mathbf{P}\)) in the same batch for efficiency, encoded by the query and passage encoders as: 
\begin{align}
    \mathcal{L}(S) &= -\frac{1}{N} \sum_i^N \log \frac{\exp(S_{(i,i)} / \tau)} {\sum_{j}^{N} \exp (S_{(i,j)} / \tau)}, \quad \text{where } S =\text{Softmax}(\mathbf{Q} \cdot \mathbf{P}^{\top}) \in \mathbb{R}^{N \times N} 
    \label{eq:infonce}
\end{align}
The in-batch negative sampling efficiently obtains \(N-1\) negative passages per query from relevant passages of other queries, as shown in Equation \ref{eq:infonce}. Consequently, the number of negative passages increases with a larger batch size. Due to this characteristic of in-batch negative sampling, dense retriever is trained using extremely large batch size, ranging from 128 to 8192 \cite{dpr, contriever, rocketqa, promptagator, spider}. However, the need to process all data in memory simultaneously requires multiple high-cost accelerators, ranging from 8 \cite{dpr, rocketqa} to 32 \cite{contriever}. This creates a hardware bottleneck that constrains various research and applications.
 
In low-resource setting, \ga\ is employed to train models with the total batch size (\(N_{\text{total}}\)), which cannot be fitted in the limited memory. \ga\ decomposes the total batch into accumulation steps, \(K\), and processes the local batch, \(N_{\text{local}} = N_\text{total} / K\), through forward and backpropagation K times to calculate gradients. The process of computing InfoNCE Loss with \ga\ is as follows.

First, the query, \(q\), and document, \(p\), are encoded by the query encoder, \(f_{\Theta}^t\), and passage encoder, \(g_{\Lambda}^t\), at training step \(t\) respectively:
\begin{equation} \label{eq:encoding}
    \mathbf{q}^t = f_{\Theta}^t(q) \in \mathbb{R}^{d_{\text{model}}}, \quad
    \mathbf{p}^t = g_{\Lambda}^t(p) \in \mathbb{R}^{d_{\text{model}}}
\end{equation}
where \(d_\text{model}\) denotes the dimension of query and passage representation. The query encoder, \(f\), and passage encoder, \(g\), are parameterized by \(\Theta\) and \(\Lambda\) respectively. The query and passage representations within the same local batch at the \(k\)-th accumulation step are given as follows:
\begin{align} \label{eq:representation set}
    \mathbf{Q}^t_k = \{\mathbf{q}^t_1, \dots ,\mathbf{q}^t_{N_{\text{local}}}\} \in \mathbb{R}^{N_\text{local} \times d_{\text{model}}},  \quad
    \mathbf{P}^t_k = \{\mathbf{p}^t_1 , \dots ,\mathbf{p}^t_{N_{\text{local}}}\} \in \mathbb{R}^{N_\text{local} \times d_{\text{model}}}
\end{align}
Using Equation \ref{eq:infonce}, the loss for the \(k\)-th accumulation step is calculated, and the loss for the total batch used for one weight update is obtained as shown in Equation \ref{eq:softmax with ga}:
\begin{align}
\label{eq:softmax with ga}
   \mathcal{L} = \frac{1}{K} \sum_{k=1}^K \mathcal{L}(S_k), \quad \text{      where  }S_k =\text{Softmax}(\mathbf{Q}^t_k \cdot (\mathbf{P}^t_k)^{\top}) \in \mathbb{R}^{N_\text{local} \times N_\text{local}}
\end{align}
In Equation \ref{eq:softmax with ga}, the number of negative passages in each accumulation step is \(N_{\text{local}} - 1\), which is fewer than the number of negative passages when using the total batch, \(N_{\text{total}} - 1\). 
This reduction in the number of negative samples results from that \ga\ use \(N_\text{local}\) passages in a single forward pass. Consequently, \ga\ cannot maintain the number of negative passages in low-resource setting, while the total amount of data used for weight updates is the same as the total batch.

% This reduction in the number of negative samples is due to \ga\ forming a smaller similarity matrix (\(S_k\)) of size (\(N_\text{local}, N_\text{local}\)) during the forward pass with the local batch. In contrast, the shape of the similarity matrix (\(N_\text{total}, N_\text{total}\)) is larger if trained in high-resourse. Consequently, the number of negative passages cannot be maintained in low-resource scenarios, while the total amount of data used for weight updates is the same with \ga\ as with the total batch.

\label{sect:proposed method}
\begin{figure}
    \centering
    \includegraphics[width=0.85\columnwidth]{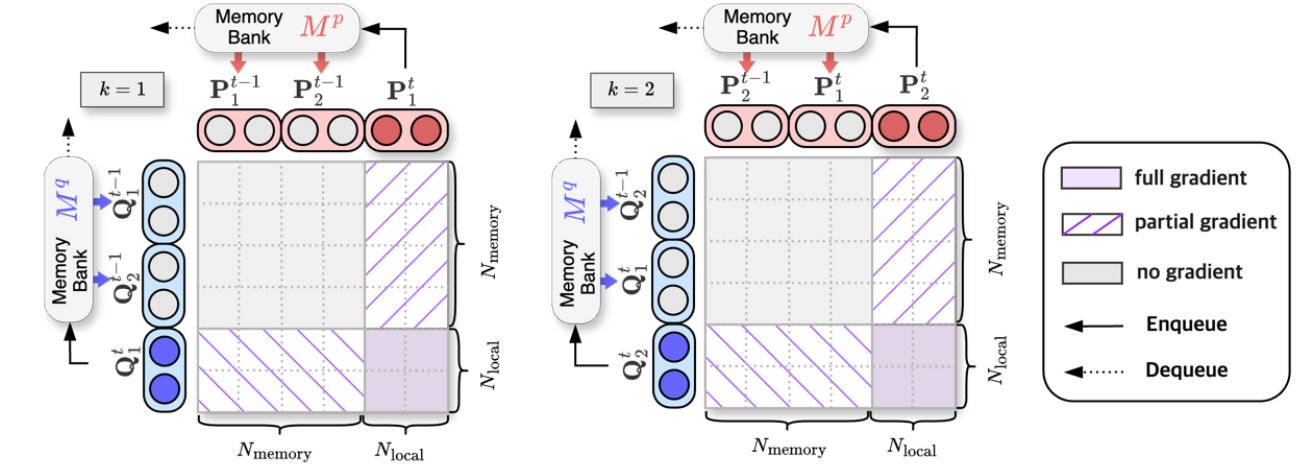}
    \caption{\textbf{Training process of \modelname\ at each accumulation step.} The illustration shows a total batch size (\(N_\text{total}\)) of 4, an accumulation step (\(K\)) of 2, and a memory bank size (\(N_\text{memory}\)) of 4. The dual memory bank caches both query and passage representations. New representations are enqueued, and the oldest are dequeued at each step, maintaining the similarity matrix (\(S_k\)) size at (\(N_\text{local} + N_\text{memory}, N_\text{local} + N_\text{memory}\)).}
    \label{fig:main-contaccum}
\end{figure}
\subsection{\modelname} 
\label{subsect:dual memory bank}
To address the issue of fewer negative passages being used with \ga, we propose \modelname, a method that utilizes a dual memory bank structure to cache representations for both queries and passages.  The query and passage memory banks (\(M_{\mathbf{q}},  M_{\mathbf{p}}\)) are implemented as First-In-First-Out queues storing \(N_{\text{memory}}^{\mathbf{q}}\) and \(N_{\text{memory}}^{\mathbf{p}}\) representations respectively. For example, as shown in Figure  \ref{fig:main-contaccum}, the oldest representations in the memory bank (\(\mathbf{P^{t-1}_1}\), \(\mathbf{Q^{t-1}_1}\)) are replaced with the newly-generated ones (\(\mathbf{P^t_1}\), \(\mathbf{Q^t_1}\)). Memory bank strategy is computationally efficient as it reuses generated representations from previous iterations \cite{cbns, xbm, densephrases}. Unlike \citet{densephrases}, which only utilized a passage memory bank \(M_\mathbf{p}\), \modelname\ employs a dual memory bank by also utilizing a query memory bank \(M_\mathbf{q}\).

\modelname\ constructs the similarity matrix using both current and stored representations from the dual memory bank as illustrated in Figure \ref{fig:main-contaccum}. It is equivalent to modifying \(S_k\) in Equation \ref{eq:softmax with ga} as: 
\begin{align} \label{eq:sim matrix with memory bank}
    \mathbf{Q} &= \mathbf{Q}^t_k \cup \text{sg}(M_{\mathbf{q}}) \in \mathbb{R}^{(N_\text{local} + N_{\text{memory}}^{\mathbf{q}}) \times d_{\text{model}}} \\   
    \mathbf{P} &= \mathbf{P}^t_k \cup \text{sg}(M_{\mathbf{p}}) \in \mathbb{R}^{(N_\text{local} + N_{\text{memory}}^{\mathbf{p}}) \times d_{\text{model}}} \\
    S_k &=\text{Softmax}(\mathbf{Q} \cdot \mathbf{P}^{\top})
\end{align}\
The backpropagation process using InfoNCE loss proceeds in the same manner as in Equation \ref{eq:softmax with ga}. However, since the representations in the memory bank do not have stored activations by the stop-gradient operation(\(\text{sg}(\cdot)\)), the gradients are not back-propagated through the representations in the memory bank.

The number of negative passages in \modelname\ is \(N_\text{local} + N_{\text{memory}}^{\mathbf{p}} - 1\), which is greater than \ga. Furthermore, if \(N_{\text{memory}}^{\mathbf{p}} > N_\text{local} \times (K-1)\),\ \modelname\ can utilize more negative passages than the total batch, enabling superior performance in low-resource setting compared to high-resource scenario.

\subsection{Gradient analysis with dual memory bank}
\label{subsect:gradient analysis}
We analyze the InfoNCE loss backpropagation process in information retrieval tasks, extending the analysis by \citet{gradcache} to consider using the memory bank. In the partial derivatives of the loss function with respect to the two encoders, \(\nabla_\Theta\mathcal{L}(S_k) = \sum_{\mathbf{q}_l \in Q_k^t} \frac{\partial \mathcal{L}(S_k)}{\partial \mathbf{q}_l} \cdot \frac{\partial \mathbf{q}_l}{\partial \Theta}, \quad \nabla_\Lambda\mathcal{L}(S_k) = \sum_{\mathbf{p}_l \in P_k^t} \frac{\partial \mathcal{L}(S_k)}{\partial \mathbf{p}_l} \cdot \frac{\partial \mathbf{p}_l}{\partial \Lambda}\), the partial derivative terms for each representation are given by:
\begin{align}
    \frac{\partial \mathcal{L}(S_k)}{\partial \mathbf{q}_l} &= -\frac{1}{N_\text{local} + N_\text{memory}^q}(\mathbf{p}_l - \sum_j^{N_\text{local} + N_\text{memory}^p }S_{k(l, j)} \cdot \mathbf{p}_j) \label{eq:derivative query}\\
    \frac{\partial \mathcal{L}(S_k)}{\partial \mathbf{p}_l} &= -\frac{1}{N_\text{local} + N_\text{memory}^q}(\mathbf{q}_l - \sum_i^{N_\text{local} + N_\text{memory}^q }S_{k(i, l)} \cdot \mathbf{q}_j), \label{eq:derivative passage}
\end{align}
where \(S_{k(i,j)}\) denotes the similarity between \(i\)-th query and \(j\)-th passage in the similarity matrix \(S_k\) of the \(k\)-th accumulation step. Detailed differentiation steps are provided in Appendix \ref{appendix:full derivatives}. 

Equations \ref{eq:derivative query} and \ref{eq:derivative passage} have a similar structure, indicating that the gradients of the two encoders are influenced by the representations generated by the opposite encoder. The difference lies in the summation targets, which are determined by the size of the memory banks. The gradient calculation for the query encoder uses \(N_\text{local} + N_{\text{memory}}^{\mathbf{p}}\) passage representations, while the passage encoder uses \(N_\text{local} + N_{\text{memory}}^{\mathbf{q}}\) query representations.

Pre-batch negatives only leverages the passage memory bank where \(N_{\text{memory}}^{\mathbf{p}} > N_{\text{memory}}^{\mathbf{q}} = 0\). The tendency where \(||\nabla_{\Theta}\mathcal{L}(S_k)||_2 < ||\nabla_{\Lambda}\mathcal{L}(S_k)||_2\) is caused by the difference in the number of representations used for the gradient calculations of the two encoders. In dual-encoder training, if the gradient norms of the two encoders remain imbalanced, the encoder with the larger gradient norm converges faster, making balanced training challenging \cite{grad-norm-imbalance, multi-task-gradient-imbalance}. Therefore, the unstable training with a memory bank is caused not only by rapid changes in encoder representations \cite{cbns, xbm}, but also by the difference in the gradient norms between the dual-encoders. We refer to this problem as the \gradimbalproblem.

The \gradimbalproblem\ can be resolved by using memory banks of equal size for queries and passages, \(N_{\text{memory}}^{\mathbf{q}} = N_{\text{memory}}^{\mathbf{p}} = N_{\text{memory}}\). This ensures that the gradient norms of the two encoders remain similar and stabilizes the training process. Further analysis is provided in Sections \ref{subsect:ablation study} and \ref{subsect:gradient norm ratio}.
%The analysis is further detailed in Sections \ref{subsect:ablation study} and \ref{subsect:gradient norm ratio}.

\section{Experimental setups}
\label{subsect:experimental setup}
\textbf{Resources}.
All experiments were conducted on a single A100 80GB GPU. For high-resource scenario, we considered situations where 80GB of memory is available. For low-resource settings, we assumed available memory as widely used commercial GPUs: 11GB (\href{https://www.nvidia.com/en-gb/geforce/graphics-cards/geforce-gtx-1080-ti/specifications/}{GTX-1080Ti}), 24GB (\href{https://www.nvidia.com/en-us/geforce/graphics-cards/30-series/rtx-3080-3080ti/}{RTX-3080Ti}, \href{https://www.nvidia.com/ko-kr/geforce/graphics-cards/40-series/rtx-4090/}{RTX-4090Ti}). To ensure strict experimental conditions, we used a function from the PyTorch \cite{pytorch} to limit the available memory.\footnote{Using the \href{https://pytorch.org/docs/stable/generated/torch.cuda.set_per_process_memory_fraction.html}{\texttt{torch.cuda.set\_per\_process\_memory\_fraction}} function in PyTorch allows for restricting the memory used during training, regardless of the total available memory.} Unless otherwise stated, all experiments assumed low resource setting where only 11GB memory is available.

\textbf{Datasets and evaluation metrics}.
The datasets used for the experiments were Natural Questions (NQ) \cite{nq}, TriviaQA \cite{triviaqa}, Curated TREC (TREC) \cite{trec}, and Web Questions (WebQ) \cite{webq} processed by DPR and MS Marco \cite{msmarco}. For Natural Questions, TriviaQA, Curated TREC, and Web Questions, we used the preprocessed data provided by DPR \cite{dpr}, which includes hard negative samples, positive passages, and answer annotations. Only queries with both positive and hard negative passages were used for training. For MS Marco, we utilized the preprocessed data from BEIR \cite{beir} and filtered BM25 \cite{bm25} hard negatives using cross-encoder scores from the sentence-transformers library \cite{setence-transformers}. Specifically, we considered passages as hard negatives if their cross-encoder scores were at least 3 points higher than the positive passages’ scores, following the preprocessing pipeline provided by sentence-transformers.

For evaluation metrics, Top@k was used for Natural Questions, TriviaQA, TREC, and WebQ following DPR. Also, we evaluate MS Marco using NDCG@K and Recall@K, widely used metrics for dense retriever. NQ and TriviaQA were evaluated using test sets, while TREC, WebQ, and MS Marco were evaluated using dev sets. Additionally, the entire document set was used for evaluation.

\textbf{Implementation details}.
The experimental code was adapted from nano-DPR\footnote{https://github.com/Hannibal046/nanoDPR}, which provides a simplified training and evaluation pipeline for DPR. All experiments were conducted using the BERT\footnote{bert-base-uncased} \cite{bert} model. To maintain consistency with DPR’s experimental setup, NQ and TREC were trained for 40 epochs, and TriviaQA and WebQ for 100 epochs. For MS Marco, performance saturated at 10 epochs, so it was trained for 10 epochs. Other training settings were also kept consistent with DPR. Detailed settings are provided in Appendix \ref{appendix: dataset and hyperparam}.
 
The optimal memory bank size, \(N_\text{memory}\), was selected using evaluation data with candidates [128, 512, 2048], resulting in 2,048 for NQ and 512 for TriviaQA. For MS Marco, WebQ, and TREC, due to the lack of evaluation data, \(N_\text{memory}\) were set based on dataset size: 1,024 for MS Marco, and 128 for WebQ and TREC.

\textbf{Baselines}.\label{subsect:baselines}
We established three baselines for each scenario, and all methods were trained with hard negatives. First, we reported the performance of DPR with the maximum batch size possible for each scenario. Further, we reported the performance of \ga\ with the total batch size of \(N_\text{total}=128\). The local batch size \(N_\text{local}\) varied by the scenario , with \(K = N_{\text{total}} / N_{\text{local}}\). We also conducted experiments with \gc\ \cite{gradcache}, known for approximating total batch performance, using the same \(N_{\text{local}}\) for single forwarding.
\section{Experimental results}  \label{sect:results}
\subsection{Performance across different resource constraints} \label{subsect:main} 
\begin{table}[h]
    \centering
    \caption{Performance of different methods in low-resource settings (11GB, 24GB) and high-resource (80GB) setting. In the high-resource setting, the score of the original DPR \cite{dpr} paper (original) and the reproduced implementation (implemented) are listed. The best score for each training environment is bolded, and scores surpassing the high-resource setting are marked with \(^{\star}\). \(N_l\) denotes the local batch size \(N_{\text{local}}\), \(N_t\) denotes the total batch size \(N_{\text{total}}\), and \(K\) represents the accumulation step.}
    \label{tab:main}
    \resizebox{\textwidth}{!}{%
    \setlength{\tabcolsep}{6pt}
    \footnotesize
        \begin{tabular}{l|c|cccccccccccc} 
        \toprule
        \multirow{3}{*}{\textbf{Method}} & {\textbf{Batch Size}}& \multicolumn{4}{c}{\textbf{MS Marco}}& \multicolumn{2}{c}{\textbf{NQ}} &\multicolumn{2}{c}{\textbf{TriviaQA}}        & \multicolumn{2}{c}{\textbf{WebQ}}            & \multicolumn{2}{c}{\textbf{TREC}}\\ 
        \cmidrule(lr){2-2}\cmidrule(lr){3-6}\cmidrule(lr){7-8}\cmidrule(lr){9-10}\cmidrule(lr){11-12}\cmidrule(lr){13-14}
                                         & \multirow{2}{*}{$N_{\text{l}}$/$K$/$N_{\text{t}}$}    & \multicolumn{2}{c}{NDCG}               & \multicolumn{2}{c}{Recall}              & \multicolumn{2}{c}{Top} & \multicolumn{2}{c}{Top} & \multicolumn{2}{c}{Top} & \multicolumn{2}{c}{Top}        \\
        \cmidrule(lr){3-4}\cmidrule(lr){5-6}\cmidrule(lr){7-8}\cmidrule(lr){9-10}\cmidrule(lr){11-12}\cmidrule(lr){13-14}
                                         & & 20 & 100 & 20 & 100 & 20 &100 &20 &100 &20 &100 &20 &100 \\
        \midrule
        \multicolumn{8}{l}{\textit{VRAM=11GB}}                                                                                                                                                                                                                        \\ 
        \midrule
        DPR        & 8/\phantom{0}1/\phantom{00}8     & 27.9\phantom{0} & 23.5\phantom{0} & 8.3\phantom{0} & 15.2\phantom{0} & 72.2\phantom{0} & 81.5\phantom{0} & 73.7\phantom{0} & 81.9\phantom{0} & 72.5\phantom{0} & 81.4\phantom{0} & 80.8\phantom{0} & 88.9\phantom{0} \\ 

        \ga\        & 8/16/128    & 31.1\phantom{0} & 26.4\phantom{0} & 10.1\phantom{0} & 18.1\phantom{0} & 77.1\phantom{0} & 84.7\phantom{0} & 78.4\phantom{0} & 84.8\phantom{0} & 74.6\phantom{0} & 81.9\phantom{0} & 79.7\phantom{0} & 89.9\phantom{0} \\
        
        \gc\        & 8/16/128    & 34.9\phantom{0} & 30.6\phantom{0} & 12.8$^\star$ & 22.4$^\star$ & 79.5$^\star$ & 85.9\phantom{0} & 79.4\phantom{0} & 85.1\phantom{0} & 75.1$^\star$ & \textbf{82.3}\phantom{0} & 81.6\phantom{0} & 90.2\phantom{0} \\
        
        \modelname\ (ours)    & 8/16/128    & \textbf{39.1}$^\star$ & \textbf{32.9}$^\star$ & \textbf{14.4}$^\star$ & \textbf{23.8}$^\star$ & \textbf{80.1}$^\star$ & \textbf{86.5}$^\star$ & \textbf{79.8}$^\star$ & \textbf{85.3}$^\star$ & \textbf{75.4}$^\star$ & 82.1\phantom{0} & \textbf{83.3}$^\star$ & \textbf{90.5}\phantom{0} \\ 
        \midrule
        
        \multicolumn{8}{l}{\textit{VRAM=24GB}} \\ 
        \midrule
        DPR        & 32/1/\phantom{0}32    & 33.1\phantom{0} & 28.6\phantom{0} & 11.5\phantom{0} & 19.6\phantom{0} & 77.0\phantom{0} & 84.8\phantom{0} & 77.5\phantom{0} & 84.2\phantom{0} & 74.8$^\star$ & 82.1\phantom{0} & \textbf{82.7}$^\star$ & \textbf{89.8}\phantom{0} \\ 
        
        \ga\        &  32/4/128 & 33.1\phantom{0} & 28.2\phantom{0} & 11.8\phantom{0} & 20.0\phantom{0} & 77.9\phantom{0} & 85.4\phantom{0} & \textbf{80.0}$^\star$ & 84.8\phantom{0} & 74.3\phantom{0} & 81.9\phantom{0} & 79.3\phantom{0} & 89.6\phantom{0} \\
        
        \gc\        & 32/4/128    & 35.5$^\star$ & 31.0$^\star$ & 12.8\phantom{0} & 22.1\phantom{0} & 79.6$^\star$ & 86.0\phantom{0} & 79.7$^\star$ & \textbf{85.1}\phantom{0} & 74.7\phantom{0} & 81.8\phantom{0} & 81.3\phantom{0} & 89.6\phantom{0} \\
        
        \modelname\ (ours)    &  32/4/128    & \textbf{39.0}$^\star$ & \textbf{32.9}$^\star$ & \textbf{14.6}$^\star$ & \textbf{24.1}$^\star$ & \textbf{80.6}$^\star$ & \textbf{86.3}$^\star$  & 79.4\phantom{0} & \textbf{85.1}\phantom{0} & \textbf{75.0}$^\star$ & \textbf{82.5}$^\star$ & 81.8\phantom{0} & 89.5\phantom{0} \\ 
        
        \midrule
        \multicolumn{8}{l}{\textit{VRAM=80GB}}\\ 
        
        \midrule
        DPR (implemented)    & 128/1/128    & 35.1\phantom{0} & 30.8\phantom{0} & 12.7\phantom{0} & 22.2\phantom{0} & 79.4\phantom{0} & 86.1\phantom{0} & 79.5\phantom{0} & 85.1\phantom{0} & 74.7\phantom{0} & 82.4\phantom{0} & 82.0\phantom{0} & 90.5\phantom{0} \\
        DPR (original)        & 128/1/128    & - & - & - & - & 78.4\phantom{0} & 85.4\phantom{0} & 79.4\phantom{0} & 85.0\phantom{0} & 73.2\phantom{0} & 81.4\phantom{0} & 79.8\phantom{0} & 89.1\phantom{0} \\

        \bottomrule
        \end{tabular}}
\end{table}
\textbf{\modelname \ outperforms the high-resource DPR even under low-resource constraints.} Table \ref{tab:main} compares the performance of \modelname with baseline methods under low-resource setting. Notably, \modelname, with only 11GB of memory, surpasses the performance of DPR in the high-resource setting (80GB). This demonstrates that \modelname is not only memory-efficient but also achieves superior performance compared to the baseline.

\textbf{\modelname \ maintains consistent performance across different memory constraints.} \modelname\ exhibits robust performance regardless of the memory constraint level (11GB or 24GB), with only minor variations between the two settings. In contrast, the performance of both DPR and \ga\ improves as the available memory increases from 11GB to 24GB. This suggests that the performance gains of \modelname\ are not significantly affected by the severity of memory limitations.

\textbf{The effectiveness of \modelname \ is amplified under more severe memory constraints.}
While \modelname\ consistently outperforms the baseline methods in both 11GB and 24GB scenarios, the performance gap between \modelname\ and the baselines is more substantial in the 11GB setting. This indicates that the advantages of \modelname\ are particularly evident when memory constraints are stringent, emphasizing its effectiveness in low-resource setting. The strong performance of \modelname\ can be attributed to its dual memory bank strategy, which allows it to utilize more negative samples than \gc, even in low-resource settings. Furthermore, \modelname\ outperforms the high-resource setting in 18 out of 24 metrics, improving up to 4.9 points. In contrast, \gc\ only surpasses the high-resource setting in 8 metrics, with marginal improvements likely due to randomness. These results demonstrate the fundamental advantage of \modelname\ in achieving superior performance compared to both the baselines and the high-resource setting.

\subsection{Influence of each components in \modelname} 
\label{subsect:ablation study} 
\begin{table}[h] 
\footnotesize
\centering
\caption{Results of removing the components of \modelname. The DPR performance in low-resource (BSZ=8) and high-resource (BSZ=128) settings are shown as baselines. The best-performing method is highlighted in bold.}
\begin{tabular}{l|c|l|c} 
\toprule
\multicolumn{2}{c|}{\textbf{w/ Hard Negative}}                                 & \multicolumn{2}{c}{\textbf{w/o Hard Negative}}                                 \\ 
\midrule
\textbf{Method}                         & \multicolumn{1}{l|}{\textbf{Top@20}} & \textbf{Method}                         & \multicolumn{1}{l}{\textbf{Top@20}}  \\ 
\midrule
DPR (BSZ=8)                             & 70.9                                & DPR (BSZ=8)                             & 63.7                                \\
DPR (BSZ=128)                           & 78.4                                & DPR (BSZ=128)                           & 74.3                                \\
\midrule
\textbf{\modelname\ (ours)}              & \textbf{78.8}                       & \textbf{\modelname\ (ours)}              & \textbf{76.3}                       \\
\multicolumn{1}{l|}{w/o. $M_q$}      & 70.8                                & \multicolumn{1}{l|}{w/o. $M_q$}      & 72.3                                \\
\multicolumn{1}{l|}{w/o. \text{Past Enc.}}       & 76.5                                & \multicolumn{1}{l|}{w/o. \text{Past Enc.}}       & 73.4                                \\
\multicolumn{1}{l|}{w/o. $M_q$/\text{Past Enc.}} & 67.8                                & \multicolumn{1}{l|}{w/o. $M_q$/\text{Past Enc.}} & 73.9                                \\
\multicolumn{1}{l|}{w/o. \ga}       & 76.7                                & \multicolumn{1}{l|}{w/o. \ga}       & 74.1\\ 
\bottomrule
\end{tabular}\label{tab:ablation}
\end{table}
Table \ref{tab:ablation} shows the influence of key components in \modelname\ by removing each component with NQ. We also reported experiments that excluded hard negatives during training to observe the tendency. The most significant performance drop occurred when the query memory bank \(M_q\) was removed, indicating its crucial role in \modelname. The other components of \modelname\ also contributed to the overall performance, with consistent trends regardless of using hard negatives.

\textbf{Passage memory bank alone degrades performance due to gradient norm imbalance.} Specifically, using only the passage memory bank (w/o. \(M_q\)), similar to the pre-batch negatives, led to an 8-point performance drop in Top@20 compared to \modelname. This decrease can be attributed to the gradient norm imbalance problem highlighted in Section \ref{subsect:gradient analysis}. Section \ref{subsect:gradient norm ratio} further analyzes this issue.

\textbf{\ga\ and past encoder representations are crucial for stable training and performance.} Moreover, when \ga\ was not applied (w/o. \ga), a 2.1-point performance decline was observed in Top@20, highlighting the importance of involving more data in gradient calculations for stable training in \modelname. Additionally, a 2.3-point performance decrease was noted when representations generated by past encoders were not used (w/o. Past Enc.). This finding confirms that past encoder representations contribute to training, as suggested by previous studies \cite{cbns, xbm, densephrases}. However, unlike pre-batch negatives, query memory bank \(M_q\) demonstrates that the greatest performance improvement is achieved by employing a dual memory bank, which leverages representations generated by past query and passage encoders.

\subsection{Memory bank size analysis} 
\label{subsect:memory-bank-analysis} 
\begin{figure}[htbp]
  \centering
  \hspace{-1.5em}
  \begin{minipage}[b]{0.6\textwidth}
    \includegraphics[width=\textwidth]{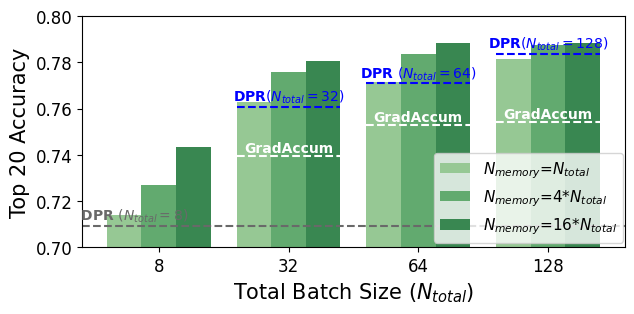}
    \caption{Analysis of accumulation step and memory bank size. DPR performance in low-resource (BSZ=8) and high-resource (BSZ=128) settings is shown as baselines, along with the performance of gradient accumulation for each total batch size (\(N_\text{total}\)).}
    \label{fig:memory-bank-analysis} 
  \end{minipage}
  \hspace{1em}
  \begin{minipage}[b]{0.35\textwidth}
    \centering
    \includegraphics[width=\textwidth]{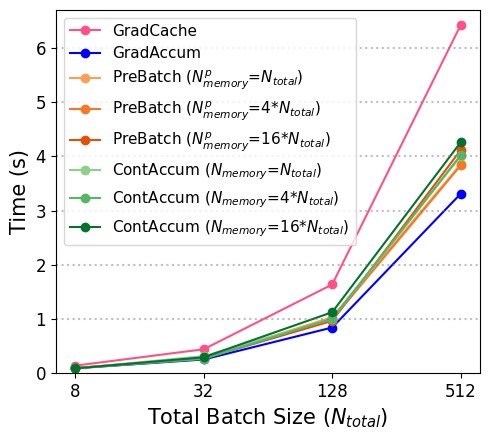}
    \caption{Comparison of the speed of one weight update for different methods as the total batch size (\(N_\text{total}\)) changes.}
    \label{fig:train speed} 
  \end{minipage}
\end{figure}
Figure \ref{fig:memory-bank-analysis} indicates the experimental results on the NQ dataset, demonstrating the impact of memory bank size \(N_\text{memory}\) and accumulation steps \(K\) on \modelname’s performance in a low-resource setting with a local batch size of 8. As the memory bank size \(N_\text{memory}\)increases, more negative passages are utilized in training, and as the accumulation steps increase, more data is considered in each model update. The performance of DPR in both low-resource and high-resource scenarios(\(N_\text{total}=32, 64, 128\)) is also included for comparison. Note that gradient accumulation is not used when the total batch size is 8 and only the dual memory bank is employed.

\textbf{\modelname\ consistently outperforms \ga\ and DPR regardless of the size of memory bank and accumulation step.} The results show that increasing the memory bank size improves performance even when \ga\ is not used. This indicates that even without gradient accumulation, utilizing representations from the memory bank to construct a larger similarity matrix enhances performance. This trend remains consistent as the accumulation step increases. Moreover, \modelname\ consistently outperforms \ga\ in all \(N_\text{total}\) settings. Remarkably, \modelname\ with \(N_\text{local}=8, N_\text{total}=64\), and \(N_\text{memory}=128\) surpasses the performance of DPR in a high-resource setting (\(N_\text{total}=N_\text{local}=128\)). The performance improvement of \modelname\ converges as the accumulation step and memory bank size increase, demonstrating that \modelname\ can robustly enhance performance regardless of memory bank size and accumulation steps.

\subsection{Train speed} 
In this subsection, we compare the training speed of \modelname\ with baseline methods. Figure \ref{fig:train speed} shows the results of experiments comparing the speed of a single training iteration (1 weight update) as the accumulation step increases in a low-resource settings with 11GB of available memory. Unlike the high-resource setting, where the total batch can be processed through forward and backward pass at once, the train speed slow down in low-resource settings due to various computations and storing gradients.

\textbf{\modelname\ achieves faster iteration times than \gc, even with large memory banks.} As shown in Figure \ref{fig:train speed}, \modelname\ performs single iterations faster than \gc\ in all total batch size. Notably, when \(N_\text{total}=512\), \gc\ is 93\% slower than \ga, while \modelname\ only takes 26\% more time, even with the largest memory bank size of \(N_\text{memory}=8192\). This indicates that \modelname\ completes iterations 34\% faster than \gc. The significant additional time for computing one iteration in \gc\ is due to the overhead of calculating and storing gradients of representations, as well as the repetitive forward and backpropagation. In contrast, \modelname\ incurs a relatively minor loss of speed compared to \ga\ due to the additional computations involved in storing and retrieving representations from the memory bank and calculating the enlarged similarity matrix. While pre-batch negatives \cite{densephrases} shows similar computational efficiency to our method, it degrades the performance as demonstrated in Table \ref{tab:ablation}.

\subsection{Gradient norm ratio} 
\label{subsect:gradient norm ratio} 
\begin{figure}[ht] 
  \centering
    \includegraphics[width=1\linewidth]{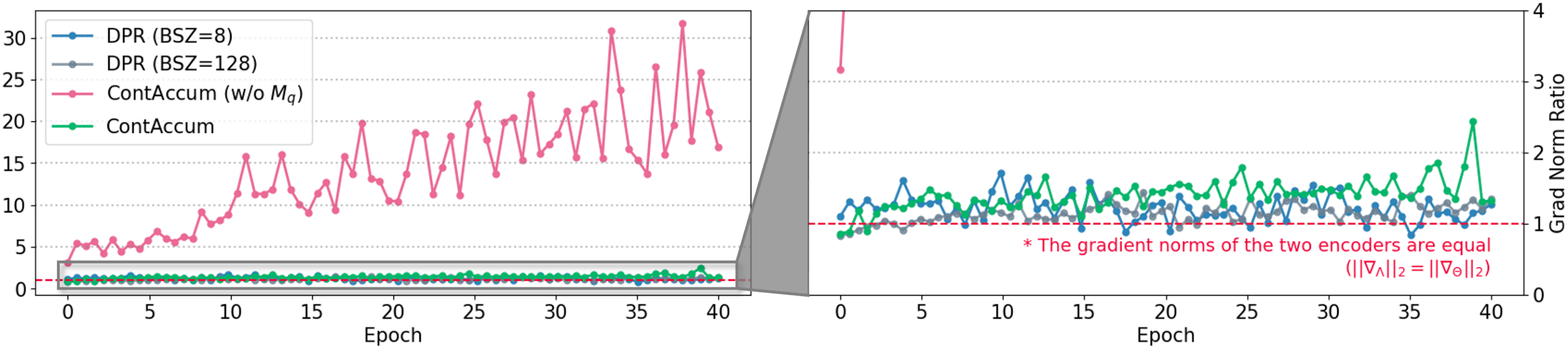}   
    \caption{Analysis of \gradnormratio\ throughout the training process on the NQ dataset.}
    \label{fig:gradnorm}
\end{figure}
We conducted experiments comparing the gradient norms of the query and passage encoders to investigate whether the presence of a query memory bank \(M_q\) affects the \gradimbalproblem, as discussed in Section \ref{subsect:gradient analysis}. The results are presented in Figure \ref{fig:gradnorm}. This experiment defines the ratio of gradient norms between the two encoders as \(\textit{GradNormRatio} = {||\nabla_\Lambda||_2 / ||\nabla\Theta||_2}\). We measured \gradnormratio\ during the training of the NQ.\footnote{The values of gradient norms are recorded after gradient clipping.} If the two encoders have similar gradient norms during training, \gradnormratio\ should be close to 1. If the passage encoder (\(g_\Lambda\)) has a larger gradient norm, \gradnormratio\ will be greater than 1.

\textbf{Dual memory bank helps maintain gradient norm balance.} The experimental results show that when the query memory bank \(M_q\) is not used, \gradnormratio\ consistently increases. In contrast, \modelname, which utilizes a dual memory bank (\(M_q, M_p\)), maintains a \gradnormratio\ close to 1, similar to DPR. 

This indicates that the pre-batch negatives exhibit \gradimbalproblem. It is because pre-batch negatives only use passage memory bank, leading to an imbalance in the number of query and passage representations used in gradient calculations, as discussed in \ref{subsect:gradient analysis}. The \gradimbalproblem\ consistently occurred even when the timing of omitting the query memory bank \(M_q\) is varied during training, as shown in Figure \ref{appendix:gradient-norm-ratio}.

The \gradimbalproblem\ observed during the actual training process becomes increasingly severe, causing the gradient norm of the passage encoder to be up to 30 times larger than the query encoder. As noted by \citet{multi-task-gradient-imbalance} and \citet{grad-norm-imbalance}, such extreme differences in gradient norms between the two models negatively impact performance. The significant performance drop observed in \ref{subsect:ablation study} when the query memory bank \(M_q\) is not used can be attributed to the \gradimbalproblem.

\section{Conclusion} 
\label{sect:conclusion}
In this work, we proposed \modelname, a novel memory reduction methodology for training dual-encoders with InfoNCE Loss in low-resource settings. By employing a dual memory bank structure, \modelname\ achieves stable training and outperforms high-resource baselines, as demonstrated through extensive experiments on five information retrieval datasets. Our mathematical analysis of the dual-encoder training process underscores the importance of balanced gradient norms, which is effectively addressed by the dual memory bank approach. Furthermore, various ablation experiments showed that the accumulation step and memory bank size significantly contribute to performance improvement.

\textbf{Limitations}. While \modelname\ reduces computational costs and stabilizes training, this study is limited by its focus on supervised fine-tuning. Recently, many studies have proposed a pre-training stage for dense retriever \cite{contriever, condenser, cocondenser, spider}. It remains to be investigated whether the \gradimbalproblem\ arises during the pre-training stage and whether \modelname\ can alleviate it. Additionally, \modelname\ still relies on the softmax operation, which incurs high computational costs. Reducing this reliance on the softmax operation could lead to more efficient training and broader application of the dense retriever.

\textbf{Broader impacts}. \modelname\ is designed to train dense retrievers efficiently, which allows it to be applied to various knowledge-intensive systems with limited resources. Examples of such applications include search engines, retrieval-augmented generation, and fact verification on local machines. However, we strongly discourage the use of \modelname\ in high-risk domains such as medical and legal fields, where the retrieval of incorrect information could have a serious impact.

\textbf{Future works}.
In future work, we plan to extend \modelname\ to the pre-training phase with a uni-encoder structure to assess its broader applicability. We also aim to investigate efficient training strategies to mitigate the substantial computational burden caused by the softmax operation. By addressing these areas, we hope to encourage further research on optimizing dual-encoder training for low-resource settings in the field of information retrieval.

\section*{Acknowledgements}
This work was supported by the National Research Foundation of Korea(NRF) grant funded by the Korea government(MSIT)(RS-2024-00407803).
This work was also supported by Institute of Information \& Communications Technology Planning \& Evaluation (IITP) grant funded by the Korea government (MSIT) (RS-2024-00460011, Climate and Environmental Data Platform for Enhancing Climate Technology Capabilities in the Anthropocene (CEDP).  \\
We would like to express our sincere gratitude to Keonwoo Kim, Joonwon Jang, Hyowon Cho, Minjin Jeon, and Sangyeop Kim for their valuable feedback and insightful comments. We also deeply appreciate our collegues; Joonghoon Kim, Saeran Park, SangMin Lee, Jiyoon Lee, Jaewon Cheon, and Seonghee Hong - for their constructive discussions and support throughout this work.

% \section*{Acknowledgements}
\newpage
\bibliographystyle{biblograph_style}
\bibliography{test}

\begin{thebibliography}{43}
\expandafter\ifx\csname natexlab\endcsname\relax\def\natexlab#1{#1}\fi

\bibitem[{Baudi{\v{s}} and {\v{S}}ediv{\'y}(2015)}]{trec}
Petr Baudi{\v{s}} and Jan {\v{S}}ediv{\'y}. 2015.
\newblock Modeling of the Question Answering Task in the YodaQA System.
\newblock In \emph{Experimental IR Meets Multilinguality, Multimodality, and Interaction}, pages 222--228, Cham. Springer International Publishing.

\bibitem[{Berant et~al.(2013)Berant, Chou, Frostig, and Liang}]{webq}
Jonathan Berant, Andrew~K. Chou, Roy Frostig, and Percy Liang. 2013.
\newblock \href {https://api.semanticscholar.org/CorpusID:6401679} {Semantic Parsing on Freebase from Question-Answer Pairs}.
\newblock In \emph{Conference on Empirical Methods in Natural Language Processing}.

\bibitem[{Chen et~al.(2020)Chen, Fan, Girshick, and He}]{mocov2}
Xinlei Chen, Haoqi Fan, Ross Girshick, and Kaiming He. 2020.
\newblock Improved Baselines with Momentum Contrastive Learning.
\newblock \emph{arXiv preprint arXiv:2003.04297}.

\bibitem[{Chen et~al.(2018)Chen, Badrinarayanan, Lee, and Rabinovich}]{grad-norm-imbalance}
Zhao Chen, Vijay Badrinarayanan, Chen-Yu Lee, and Andrew Rabinovich. 2018.
\newblock \href {https://proceedings.mlr.press/v80/chen18a.html} {{G}rad{N}orm: Gradient Normalization for Adaptive Loss Balancing in Deep Multitask Networks}.
\newblock In \emph{Proceedings of the 35th International Conference on Machine Learning}, volume~80 of \emph{Proceedings of Machine Learning Research}, pages 794--803. PMLR.

\bibitem[{Dai et~al.(2023)Dai, Zhao, Ma, Luan, Ni, Lu, Bakalov, Guu, Hall, and Chang}]{promptagator}
Zhuyun Dai, Vincent~Y Zhao, Ji~Ma, Yi~Luan, Jianmo Ni, Jing Lu, Anton Bakalov, Kelvin Guu, Keith Hall, and Ming-Wei Chang. 2023.
\newblock \href {https://openreview.net/forum?id=gmL46YMpu2J} {Promptagator: Few-shot Dense Retrieval From 8 Examples}.
\newblock In \emph{The Eleventh International Conference on Learning Representations}.

\bibitem[{Devlin et~al.(2019)Devlin, Chang, Lee, and Toutanova}]{bert}
Jacob Devlin, Ming-Wei Chang, Kenton Lee, and Kristina Toutanova. 2019.
\newblock \href {https://doi.org/10.18653/v1/N19-1423} {{BERT}: Pre-training of Deep Bidirectional Transformers for Language Understanding}.
\newblock In \emph{Proceedings of the 2019 Conference of the North {A}merican Chapter of the Association for Computational Linguistics: Human Language Technologies, Volume 1 (Long and Short Papers)}, pages 4171--4186, Minneapolis, Minnesota. Association for Computational Linguistics.

\bibitem[{Gao and Callan(2021)}]{condenser}
Luyu Gao and Jamie Callan. 2021.
\newblock \href {https://doi.org/10.18653/v1/2021.emnlp-main.75} {Condenser: a Pre-training Architecture for Dense Retrieval}.
\newblock In \emph{Proceedings of the 2021 Conference on Empirical Methods in Natural Language Processing}, pages 981--993, Online and Punta Cana, Dominican Republic. Association for Computational Linguistics.

\bibitem[{Gao and Callan(2022)}]{cocondenser}
Luyu Gao and Jamie Callan. 2022.
\newblock \href {https://doi.org/10.18653/v1/2022.acl-long.203} {Unsupervised Corpus Aware Language Model Pre-training for Dense Passage Retrieval}.
\newblock In \emph{Proceedings of the 60th Annual Meeting of the Association for Computational Linguistics (Volume 1: Long Papers)}, pages 2843--2853, Dublin, Ireland. Association for Computational Linguistics.

\bibitem[{Gao et~al.(2021)Gao, Zhang, Han, and Callan}]{gradcache}
Luyu Gao, Yunyi Zhang, Jiawei Han, and Jamie Callan. 2021.
\newblock Scaling Deep Contrastive Learning Batch Size under Memory Limited Setup.
\newblock In \emph{Proceedings of the 6th Workshop on Representation Learning for NLP}.

\bibitem[{Gutmann and Hyvärinen(2010)}]{nce}
Michael Gutmann and Aapo Hyvärinen. 2010.
\newblock \href {https://proceedings.mlr.press/v9/gutmann10a.html} {Noise-contrastive estimation: A new estimation principle for unnormalized statistical models}.
\newblock In \emph{Proceedings of the Thirteenth International Conference on Artificial Intelligence and Statistics}, volume~9 of \emph{Proceedings of Machine Learning Research}, pages 297--304, Chia Laguna Resort, Sardinia, Italy. PMLR.

\bibitem[{He et~al.(2020)He, Fan, Wu, Xie, and Girshick}]{mocov1}
K.~He, H.~Fan, Y.~Wu, S.~Xie, and R.~Girshick. 2020.
\newblock \href {https://doi.org/10.1109/CVPR42600.2020.00975} {Momentum Contrast for Unsupervised Visual Representation Learning}.
\newblock In \emph{2020 IEEE/CVF Conference on Computer Vision and Pattern Recognition (CVPR)}, pages 9726--9735, Los Alamitos, CA, USA. IEEE Computer Society.

\bibitem[{Izacard et~al.(2022)Izacard, Caron, Hosseini, Riedel, Bojanowski, Joulin, and Grave}]{contriever}
Gautier Izacard, Mathilde Caron, Lucas Hosseini, Sebastian Riedel, Piotr Bojanowski, Armand Joulin, and Edouard Grave. 2022.
\newblock \href {https://openreview.net/forum?id=jKN1pXi7b0} {Unsupervised Dense Information Retrieval with Contrastive Learning}.
\newblock \emph{Transactions on Machine Learning Research}.

\bibitem[{Jin et~al.(2023)Jin, Zeng, Wang, Chen, Wei, Li, Wang, Li, Li, Lu, Wang, Han, and Tang}]{lm-indexer}
Bowen Jin, Hansi Zeng, Guoyin Wang, Xiusi Chen, Tianxin Wei, Ruirui Li, Zhengyang Wang, Zheng Li, Yang Li, Hanqing Lu, Suhang Wang, Jiawei Han, and Xianfeng Tang. 2023.
\newblock \href {http://arxiv.org/abs/2310.07815} {Language Models As Semantic Indexers}.

\bibitem[{Johnson et~al.(2021)Johnson, Douze, and Jégou}]{faiss}
Jeff Johnson, Matthijs Douze, and Hervé Jégou. 2021.
\newblock \href {https://doi.org/10.1109/TBDATA.2019.2921572} {Billion-Scale Similarity Search with GPUs}.
\newblock \emph{IEEE Transactions on Big Data}, 7(3):535--547.

\bibitem[{Joshi et~al.(2017)Joshi, Choi, Weld, and Zettlemoyer}]{triviaqa}
Mandar Joshi, Eunsol Choi, Daniel~S. Weld, and Luke Zettlemoyer. 2017.
\newblock TriviaQA: A Large Scale Distantly Supervised Challenge Dataset for Reading Comprehension.
\newblock In \emph{Proceedings of the 55th Annual Meeting of the Association for Computational Linguistics}, Vancouver, Canada. Association for Computational Linguistics.

\bibitem[{Karpukhin et~al.(2020)Karpukhin, Oguz, Min, Lewis, Wu, Edunov, Chen, and Yih}]{dpr}
Vladimir Karpukhin, Barlas Oguz, Sewon Min, Patrick Lewis, Ledell Wu, Sergey Edunov, Danqi Chen, and Wen-tau Yih. 2020.
\newblock \href {https://doi.org/10.18653/v1/2020.emnlp-main.550} {Dense Passage Retrieval for Open-Domain Question Answering}.
\newblock In \emph{Proceedings of the 2020 Conference on Empirical Methods in Natural Language Processing (EMNLP)}, pages 6769--6781, Online. Association for Computational Linguistics.

\bibitem[{Kulis(2013)}]{metric-learning}
Brian Kulis. 2013.
\newblock Metric learning: A survey.
\newblock \emph{Foundations and Trends in Machine Learning}, 5(4):287--364.

\bibitem[{Kwiatkowski et~al.(2019)Kwiatkowski, Palomaki, Redfield, Collins, Parikh, Alberti, Epstein, Polosukhin, Kelcey, Devlin, Lee, Toutanova, Jones, Chang, Dai, Uszkoreit, Le, and Petrov}]{nq}
Tom Kwiatkowski, Jennimaria Palomaki, Olivia Redfield, Michael Collins, Ankur Parikh, Chris Alberti, Danielle Epstein, Illia Polosukhin, Matthew Kelcey, Jacob Devlin, Kenton Lee, Kristina~N. Toutanova, Llion Jones, Ming-Wei Chang, Andrew Dai, Jakob Uszkoreit, Quoc Le, and Slav Petrov. 2019.
\newblock Natural Questions: a Benchmark for Question Answering Research.
\newblock \emph{Transactions of the Association of Computational Linguistics}.

\bibitem[{Lee et~al.(2021)Lee, Sung, Kang, and Chen}]{densephrases}
Jinhyuk Lee, Mujeen Sung, Jaewoo Kang, and Danqi Chen. 2021.
\newblock Learning Dense Representations of Phrases at Scale.
\newblock In \emph{Association for Computational Linguistics (ACL)}.

\bibitem[{Lee et~al.(2019)Lee, Chang, and Toutanova}]{orqa}
Kenton Lee, Ming-Wei Chang, and Kristina Toutanova. 2019.
\newblock \href {https://doi.org/10.18653/v1/P19-1612} {Latent Retrieval for Weakly Supervised Open Domain Question Answering}.
\newblock In \emph{Proceedings of the 57th Annual Meeting of the Association for Computational Linguistics}, pages 6086--6096, Florence, Italy. Association for Computational Linguistics.

\bibitem[{Li et~al.(2022)Li, Li, Xiong, and Hoi}]{blip}
Junnan Li, Dongxu Li, Caiming Xiong, and Steven C.~H. Hoi. 2022.
\newblock \href {https://api.semanticscholar.org/CorpusID:246411402} {BLIP: Bootstrapping Language-Image Pre-training for Unified Vision-Language Understanding and Generation}.
\newblock In \emph{International Conference on Machine Learning}.

\bibitem[{Li et~al.(2021)Li, Selvaraju, Gotmare, Joty, Xiong, and Hoi}]{albef}
Junnan Li, Ramprasaath Selvaraju, Akhilesh Gotmare, Shafiq Joty, Caiming Xiong, and Steven Chu~Hong Hoi. 2021.
\newblock \href {https://proceedings.neurips.cc/paper_files/paper/2021/file/505259756244493872b7709a8a01b536-Paper.pdf} {Align before Fuse: Vision and Language Representation Learning with Momentum Distillation}.
\newblock In \emph{Advances in Neural Information Processing Systems}, volume~34, pages 9694--9705. Curran Associates, Inc.

\bibitem[{Loshchilov and Hutter(2019)}]{adamw}
Ilya Loshchilov and Frank Hutter. 2019.
\newblock \href {https://openreview.net/forum?id=Bkg6RiCqY7} {Decoupled Weight Decay Regularization}.
\newblock In \emph{International Conference on Learning Representations}.

\bibitem[{Lu et~al.(2021)Lu, He, Xiong, Ke, Malik, Dou, Bennett, Liu, and Overwijk}]{seed}
Shuqi Lu, Di~He, Chenyan Xiong, Guolin Ke, Waleed Malik, Zhicheng Dou, Paul Bennett, Tie-Yan Liu, and Arnold Overwijk. 2021.
\newblock \href {https://doi.org/10.18653/v1/2021.emnlp-main.220} {Less is More: Pretrain a Strong {S}iamese Encoder for Dense Text Retrieval Using a Weak Decoder}.
\newblock In \emph{Proceedings of the 2021 Conference on Empirical Methods in Natural Language Processing}, pages 2780--2791, Online and Punta Cana, Dominican Republic. Association for Computational Linguistics.

\bibitem[{Min et~al.(2023)Min, Shi, Lewis, Chen, Yih, Hajishirzi, and Zettlemoyer}]{npm}
Sewon Min, Weijia Shi, Mike Lewis, Xilun Chen, Wen-tau Yih, Hannaneh Hajishirzi, and Luke Zettlemoyer. 2023.
\newblock \href {https://doi.org/10.18653/v1/2023.findings-acl.132} {Nonparametric Masked Language Modeling}.
\newblock In \emph{Findings of the Association for Computational Linguistics: ACL 2023}, pages 2097--2118, Toronto, Canada. Association for Computational Linguistics.

\bibitem[{Nguyen et~al.(2016)Nguyen, Rosenberg, Song, Gao, Tiwary, Majumder, and Deng}]{msmarco}
Tri Nguyen, Mir Rosenberg, Xia Song, Jianfeng Gao, Saurabh Tiwary, Rangan Majumder, and Li~Deng. 2016.
\newblock MS MARCO: A human generated machine reading comprehension dataset.
\newblock In \emph{CoCo@ NIPS}.

\bibitem[{Paszke et~al.(2019)Paszke, Gross, Massa, Lerer, Bradbury, Chanan, Killeen, Lin, Gimelshein, Antiga, Desmaison, Kopf, Yang, DeVito, Raison, Tejani, Chilamkurthy, Steiner, Fang, Bai, and Chintala}]{pytorch}
Adam Paszke, Sam Gross, Francisco Massa, Adam Lerer, James Bradbury, Gregory Chanan, Trevor Killeen, Zeming Lin, Natalia Gimelshein, Luca Antiga, Alban Desmaison, Andreas Kopf, Edward Yang, Zachary DeVito, Martin Raison, Alykhan Tejani, Sasank Chilamkurthy, Benoit Steiner, Lu~Fang, Junjie Bai, and Soumith Chintala. 2019.
\newblock \href {http://papers.neurips.cc/paper/9015-pytorch-an-imperative-style-high-performance-deep-learning-library.pdf} {PyTorch: An Imperative Style, High-Performance Deep Learning Library}.
\newblock In \emph{Advances in Neural Information Processing Systems 32}, pages 8024--8035. Curran Associates, Inc.

\bibitem[{Qu et~al.(2021)Qu, Ding, Liu, Liu, Ren, Zhao, Dong, Wu, and Wang}]{rocketqa}
Yingqi Qu, Yuchen Ding, Jing Liu, Kai Liu, Ruiyang Ren, Wayne~Xin Zhao, Daxiang Dong, Hua Wu, and Haifeng Wang. 2021.
\newblock \href {https://doi.org/10.18653/v1/2021.naacl-main.466} {{R}ocket{QA}: An Optimized Training Approach to Dense Passage Retrieval for Open-Domain Question Answering}.
\newblock In \emph{Proceedings of the 2021 Conference of the North American Chapter of the Association for Computational Linguistics: Human Language Technologies}, pages 5835--5847, Online. Association for Computational Linguistics.

\bibitem[{Ram et~al.(2022)Ram, Shachaf, Levy, Berant, and Globerson}]{spider}
Ori Ram, Gal Shachaf, Omer Levy, Jonathan Berant, and Amir Globerson. 2022.
\newblock \href {https://doi.org/10.18653/v1/2022.naacl-main.193} {Learning to Retrieve Passages without Supervision}.
\newblock In \emph{Proceedings of the 2022 Conference of the North American Chapter of the Association for Computational Linguistics: Human Language Technologies}, pages 2687--2700, Seattle, United States. Association for Computational Linguistics.

\bibitem[{Reimers and Gurevych(2019)}]{setence-transformers}
Nils Reimers and Iryna Gurevych. 2019.
\newblock \href {https://arxiv.org/abs/1908.10084} {Sentence-BERT: Sentence Embeddings using Siamese BERT-Networks}.
\newblock In \emph{Proceedings of the 2019 Conference on Empirical Methods in Natural Language Processing}. Association for Computational Linguistics.

\bibitem[{Ren et~al.(2021)Ren, Lv, Qu, Liu, Zhao, She, Wu, Wang, and Wen}]{pair}
Ruiyang Ren, Shangwen Lv, Yingqi Qu, Jing Liu, Wayne~Xin Zhao, Qiaoqiao She, Hua Wu, Haifeng Wang, and Ji{-}Rong Wen. 2021.
\newblock \href {https://doi.org/10.18653/V1/2021.FINDINGS-ACL.191} {{PAIR:} Leveraging Passage-Centric Similarity Relation for Improving Dense Passage Retrieval}.
\newblock In \emph{Findings of the Association for Computational Linguistics: {ACL/IJCNLP} 2021, Online Event, August 1-6, 2021}, volume {ACL/IJCNLP} 2021 of \emph{Findings of {ACL}}, pages 2173--2183. Association for Computational Linguistics.

\bibitem[{Robertson and Zaragoza(2009)}]{bm25}
Stephen Robertson and Hugo Zaragoza. 2009.
\newblock \href {https://doi.org/10.1561/1500000019} {The Probabilistic Relevance Framework: BM25 and Beyond}.
\newblock \emph{Foundations and Trends in Information Retrieval}, 3:333--389.

\bibitem[{Senushkin et~al.(2023)Senushkin, Patakin, Kuznetsov, and Konushin}]{multi-task-gradient-imbalance}
Dmitry Senushkin, Nikolay Patakin, Arseny Kuznetsov, and Anton Konushin. 2023.
\newblock Independent Component Alignment for Multi-Task Learning.
\newblock In \emph{Proceedings of the IEEE/CVF Conference on Computer Vision and Pattern Recognition (CVPR)}, pages 20083--20093.

\bibitem[{Sun et~al.(2023)Sun, Yan, Chen, Wang, Zhu, Ren, Chen, Yin, Rijke, and Ren}]{genret}
Weiwei Sun, Lingyong Yan, Zheng Chen, Shuaiqiang Wang, Haichao Zhu, Pengjie Ren, Zhumin Chen, Dawei Yin, Maarten Rijke, and Zhaochun Ren. 2023.
\newblock \href {https://proceedings.neurips.cc/paper_files/paper/2023/file/91228b942a4528cdae031c1b68b127e8-Paper-Conference.pdf} {Learning to Tokenize for Generative Retrieval}.
\newblock In \emph{Advances in Neural Information Processing Systems}, volume~36, pages 46345--46361. Curran Associates, Inc.

\bibitem[{Thakur et~al.(2021)Thakur, Reimers, R\"{u}ckl\'{e}, Srivastava, and Gurevych}]{beir}
Nandan Thakur, Nils Reimers, Andreas R\"{u}ckl\'{e}, Abhishek Srivastava, and Iryna Gurevych. 2021.
\newblock \href {https://datasets-benchmarks-proceedings.neurips.cc/paper_files/paper/2021/file/65b9eea6e1cc6bb9f0cd2a47751a186f-Paper-round2.pdf} {BEIR: A Heterogeneous Benchmark for Zero-shot Evaluation of Information Retrieval Models}.
\newblock In \emph{Proceedings of the Neural Information Processing Systems Track on Datasets and Benchmarks}, volume~1.

\bibitem[{van~den Oord et~al.(2018)van~den Oord, Li, and Vinyals}]{code}
A{\"{a}}ron van~den Oord, Yazhe Li, and Oriol Vinyals. 2018.
\newblock \href {http://arxiv.org/abs/1807.03748} {Representation Learning with Contrastive Predictive Coding}.
\newblock \emph{CoRR}, abs/1807.03748.

\bibitem[{Wang et~al.(2021)Wang, Zhu, and He}]{cbns}
Jinpeng Wang, Jieming Zhu, and Xiuqiang He. 2021.
\newblock \href {https://doi.org/10.1145/3404835.3463032} {Cross-Batch Negative Sampling for Training Two-Tower Recommenders}.
\newblock In \emph{Proceedings of the 44th International ACM SIGIR Conference on Research and Development in Information Retrieval}, SIGIR '21, page 1632–1636, New York, NY, USA. Association for Computing Machinery.

\bibitem[{Wang et~al.(2020)Wang, Zhang, Huang, and Scott}]{xbm}
Xun Wang, Haozhi Zhang, Weilin Huang, and Matthew~R Scott. 2020.
\newblock Cross-Batch Memory for Embedding Learning.
\newblock In \emph{CVPR}.

\bibitem[{Wu et~al.(2018{\natexlab{a}})Wu, Xiong, Yu, and Lin}]{memory-bank-neg}
Z.~Wu, Y.~Xiong, S.~X. Yu, and D.~Lin. 2018{\natexlab{a}}.
\newblock \href {https://doi.org/10.1109/CVPR.2018.00393} {Unsupervised Feature Learning via Non-parametric Instance Discrimination}.
\newblock In \emph{2018 IEEE/CVF Conference on Computer Vision and Pattern Recognition (CVPR)}, pages 3733--3742, Los Alamitos, CA, USA. IEEE Computer Society.

\bibitem[{Wu et~al.(2018{\natexlab{b}})Wu, Efros, and Yu}]{memory-bank}
Zhirong Wu, Alexei~A. Efros, and Stella~X. Yu. 2018{\natexlab{b}}.
\newblock Improving Generalization via Scalable Neighborhood Component Analysis.
\newblock In \emph{Proceedings of the European Conference on Computer Vision (ECCV)}.

\bibitem[{Xiong et~al.(2021)Xiong, Xiong, Li, Tang, Liu, Bennett, Ahmed, and Overwijk}]{ance}
Lee Xiong, Chenyan Xiong, Ye~Li, Kwok-Fung Tang, Jialin Liu, Paul~N. Bennett, Junaid Ahmed, and Arnold Overwijk. 2021.
\newblock \href {https://openreview.net/forum?id=zeFrfgyZln} {Approximate Nearest Neighbor Negative Contrastive Learning for Dense Text Retrieval}.
\newblock In \emph{International Conference on Learning Representations}.

\bibitem[{Yang et~al.(2024)Yang, Shao, Dong, and Tang}]{trisampler}
Zhen Yang, Zhou Shao, Yuxiao Dong, and Jie Tang. 2024.
\newblock \href {https://doi.org/10.1609/aaai.v38i8.28779} {TriSampler: A Better Negative Sampling Principle for Dense Retrieval}.
\newblock \emph{Proceedings of the AAAI Conference on Artificial Intelligence}, 38(8):9269--9277.

\bibitem[{Zhao et~al.(2024)Zhao, Liu, Ren, and Wen}]{retrieval-survey}
Wayne~Xin Zhao, Jing Liu, Ruiyang Ren, and Ji-Rong Wen. 2024.
\newblock \href {https://doi.org/10.1145/3637870} {Dense Text Retrieval Based on Pretrained Language Models: A Survey}.
\newblock \emph{ACM Trans. Inf. Syst.}, 42(4).

\end{thebibliography}

\newpage
\section*{A. Derivatives of InfoLoss with memory bank} \label{appendix:full derivatives}
\begin{align*} 
    \nabla_\Theta\mathcal{L}(S_k) &= \sum_{\mathbf{q}_l \in Q_k^t} = \frac{\partial \mathcal{L}(S_k)}{\partial \mathbf{q}_l} \cdot \frac{\partial \mathbf{q}_l}{\partial \Theta} \\
    \frac{\partial \mathcal{L}(S_k)}{\partial \mathbf{q}_l} &= \frac{\partial}{\partial \mathbf{q}_l}(-\frac{1}{N_\text{local} + N_\text{memory}^q}) \sum_i^{N_\text{local} + N_\text{memory}^q} \log \frac{\exp(\mathbf{q}_i \cdot \mathbf{p}_i^\top)}{\sum_j^{N_\text{local} + N_\text{memory}^I p} \exp(\mathbf{q}_i \cdot \mathbf{p}_j^\top)} \\
        &= -\frac{1}{N_\text{local} + N_\text{memory}^q} \sum_i^{N_\text{local} + N_\text{memory}^q} \left[\frac{\partial}{\partial \mathbf{q}_l}(\mathbf{q}_i \cdot \mathbf{p}_i^\top) - \frac{\partial}{\partial \mathbf{q}_l}\log \sum_j^{N_\text{local} + N_\text{memory}^p} \exp(\mathbf{q}_i \cdot \mathbf{p}_j^\top)\right] \\
        &= -\frac{1}{N_\text{local} + N_\text{memory}^q}\left(\mathbf{p}_l - \sum_i^{N_\text{local} + N_{\text{memory}}^q}\sum_j^{N_\text{local} + N_{\text{memory}}^p}\left[\frac{\exp (\mathbf{q}_i \cdot \mathbf{p}_j^\top)}{\sum_k^{N_{\text{local}} + N_{\text{memory}}^p}\exp(\mathbf{q}_i \cdot \mathbf{p}_k^\top)} \cdot \frac{\partial}{\partial \mathbf{q}_j}\left(\mathbf{q}_i \cdot \mathbf{p}_j^\top\right)\right]\right)\\
        &= -\frac{1}{N_\text{local} + N_\text{memory}^q}\left(\mathbf{p}_l - \sum_i^{N_\text{local} + N_\text{memory}^q} \sum_j^{N_\text{local} + N_\text{memory}^p}\left[S_{k(i,j)} \cdot \frac{\partial}{\partial \mathbf{q}_l}(\mathbf{q}_i \cdot \mathbf{p}_j^\top)\right]\right) \\
        &= -\frac{1}{N_\text{local} + N_\text{memory}^q}\left(\mathbf{p}_l - \sum_j^{N_\text{local} + N_\text{memory}^p} \left[S_{k(l, j)} \cdot \mathbf{p}_j\right]\right) \\
    \nabla_\Lambda\mathcal{L}(S_k) &= \sum_{\textbf{p}_l in P_k^t} = \frac{\partial \mathcal{L}(S_k)}{\partial \mathbf{p}_l} \cdot \frac{\partial \mathbf{p}_l}{\partial \Lambda} \\
    \frac{\partial \mathcal{L}(S_k)}{\partial \mathbf{p}_l} &=  \frac{\partial}{\partial \mathbf{p}_l}(-\frac{1}{N_\text{local} + N_\text{memory}^q} \sum_i^{N_\text{local} + N_\text{memory}^q} \log \frac{\exp (\mathbf{q}_i \cdot \mathbf{p}_i^\top)}{\sum_j^{N_\text{local} + N_\text{memory}^p} \exp (\mathbf{q}_i \cdot \mathbf{p}_j^\top)} \\
        &= -\frac{1}{N_\text{local} + N_\text{memory}^q}\sum_i^{N_\text{local} + N_\text{memory}^q} \left[\frac{\partial}{\partial \mathbf{p}_l} (\mathbf{q}_i \cdot \mathbf{p}_i^\top)  - \frac{\partial}{\partial \mathbf{p}_l} \log \sum_j^{N_\text{local} + N_\text{memory}^p} \exp(\mathbf{q}_i \cdot \mathbf{p}_j^\top)\right]\\
        &= -\frac{1}{N_\text{local} + N_\text{memory}^q} \left(\mathbf{q}_l - \sum_i^{N_\text{local} + N_\text{memory}^q} \sum_j^{N_\text{local} + N_\text{memory}^p} \left[\frac{\exp (\mathbf{q}_i \cdot \mathbf{p}_j^\top)}{\sum_k^{N_\text{local} + N_\text{memory}^p} \exp(\mathbf{q}_i \cdot \mathbf{p}_k^\top)} \cdot \frac{\partial}{\partial \mathbf{p}_l(\mathbf{q}_i \cdot \mathbf{p}_j^\top)} \right] \right) \\
        &= -\frac{1}{N_\text{local} + N_\text{memory}^q}\left(\mathbf{q}_l - \sum_i^{N_\text{local} + N_\text{memory}^q} \sum_j^{N_\text{local} + N_\text{memory}^p} \left[S_{k(i, j)} \cdot \frac{\partial}{\partial \mathbf{p}_l}(\mathbf{q}_i \cdot \mathbf{p}_j^\top) \right] \right)\\
        &= -\frac{1}{N_\text{local} + N_\text{memory}^q}\left(\mathbf{q}_l - \sum_i^{N_\text{local} + N_\text{memory}^q} S_{k(i, l)} \cdot \mathbf{q}_l \right)
\end{align*}

\section*{B. Details on hyperparameters} 
\label{appendix: dataset and hyperparam}
% \textbf{Datasets}. For Natural Questions, TriviaQA, Curated TREC, and Web Questions, we used the preprocessed data provided by DPR \cite{dpr}, which includes hard negative samples, positive passages, and answer annotations. Only queries with both positive and hard negative passages were used for training. For MS Marco, we utilized the preprocessed data from BEIR \cite{beir} and filtered BM25 hard negatives using cross-encoder scores from the sentence-transformers library. Specifically, we considered passages as hard negatives if their cross-encoder scores were at least 3 points higher than the positive passages’ scores, following the preprocessing pipeline provided by sentence-transformers \cite{setence-transformers}.

\textbf{Hyperparameters}. The hyperparameters for training were set as follows: the warmup step was 1,237 steps, weight decay was set to 0, and a customized scheduler with a linear decay of the learning rate after the warmup was used. The optimizer was AdamW \cite{adamw} with epsilon set to 1e-8, and the learning rate was 2e-5. Gradient clipping was applied at a value of 2.0, and \(\tau\) was set to 1. For retrieval, we used the FAISS \cite{faiss} library to perform exact nearest neighbor search with default hyperparameters.

\section*{C. Similarity Mass} \label{appendix:prob mass} 
\begin{figure}[ht] 
  \centering
  \caption{Experiments on similarity probability mass.}
    \includegraphics[width=0.8\linewidth]{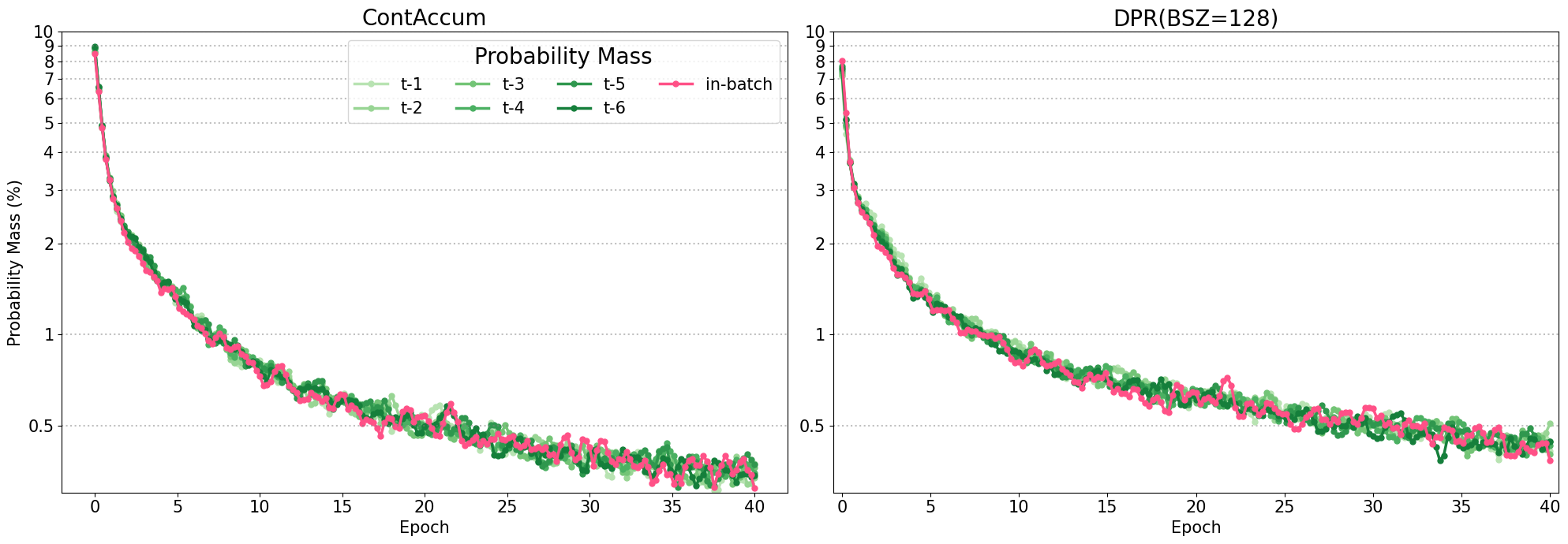}
    \label{fig:probmass figure}
\end{figure}
To verify whether representations generated by past encoders aid the current encoder’s training, we conducted an experiment measuring the similarity mass of passage representations at different time steps. The results are shown in Figure \ref{fig:probmass figure}. The similarity mass is defined as the sum of similarities after passing through a softmax function for all current time \(t\) queries with passage representations generated at past time steps \(t-k\), as shown in Equation \ref{eq:similarity mass}:
\begin{align} \label{eq:similarity mass}
    \text{SimMass}_{t-k} = \frac{1}{|\mathbf{Q}^t|} \sum^{|\mathbf{Q}^t|}_{i=1} \sum^{|\mathbf{P}^{t-k}|}_{j=1}\mathbf{Q}^t_i \cdot (\mathbf{P}^{t-k}_j)^\top
\end{align}
\textbf{Passage representations of the current and previous encoder have similar importance as negative passage.}
The results indicate that there is no significant difference in the similarity mass between the in-batch negative passage representations at the current training step and the passage representations from up to six previous steps. As shown in Equation \ref{eq:derivative query} and \ref{eq:derivative passage}, the gradients of the two encoders are proportional to the magnitude of the similarities. This means that negative passages with high similarity to a single query produce large gradients, which aids in training the dense retrieval model \cite{ance}. This finding suggests that past representations can be beneficial from the early stages of training, contrary to previous studies \cite{cbns, xbm}.

Additionally, as illustrated in Figure \ref{fig:probmass figure}, \modelname\ demonstrates the same similarity mass trend as DPR, validating the effectiveness of utilizing past representations from the early stages of training with \modelname.

\section*{D. Gradient Norm Ratio of Omitting the Query Memory Bank} 
\label{appendix:gradient-norm-ratio}
%여기에 넣깅
\begin{figure}[ht] 
  \centering
  \caption{Experimental results of omitting query memory bank during training.}
    \includegraphics[width=0.8\linewidth]{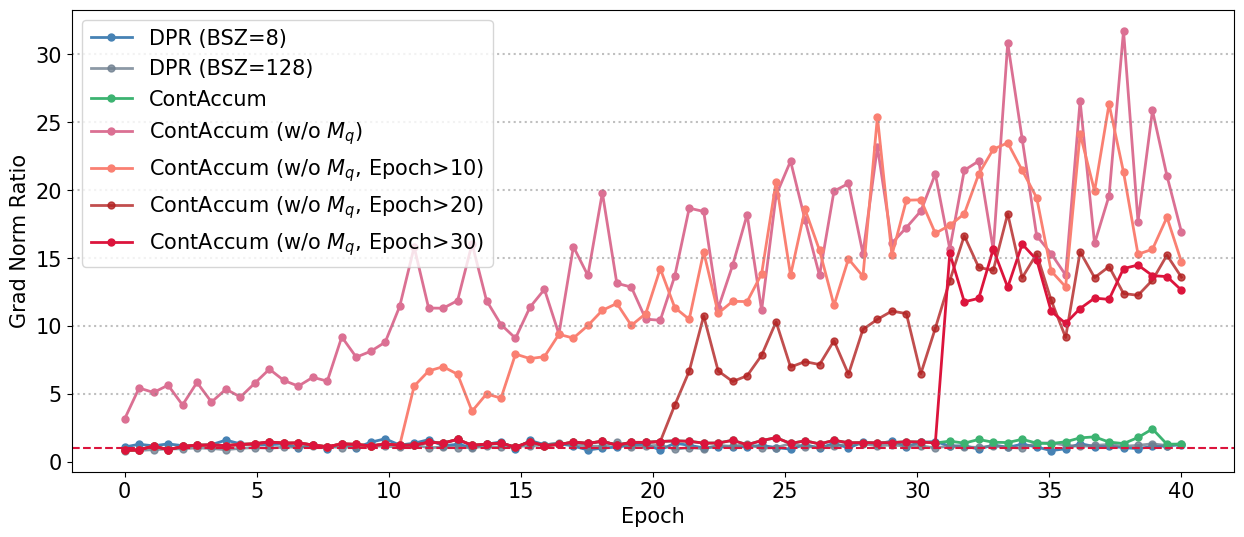}
    \label{fig:gradient-norm-ratio-appendix}
\end{figure}
 \textbf{\textit{Gradient norm imbalance problem} occurs when the query memory bank is omitted}. We omitted the query memory bank during training at various epochs: [10, 20, 30]. As shown in Figure \ref{fig:gradient-norm-ratio-appendix}, the gradient norm imbalance problem arises immediately after the query memory bank is excluded. Additionally, irrespective of when the query memory bank is omitted, all experiments without the query memory bank exhibit very high gradient norm ratios in the later stages of training. This indicates that \gradimbalproblem\ can cause unstable training during the entire training process, unlike previous studies which mentioned the major cause of unstable training is rapid changes in encoder representations in the early epochs \cite{xbm, cbns}.

 \section*{E. Actual Memory Usage}
\textbf{ContAccum uses few memory for dual memory bank but it works greatly}
Theoretically, \modelname's query and passage memory banks(\(M_q, M_p\)) do not cache activation values, requiring only additional memory for the stored representations compared to \ga. The memory usage of the dual memory bank can be calculated as follows: 
\begin{equation}
    \textbf{N}_\text{memory} \times \text{dim}_\text{embed} \times 2 \times 4
\end{equation}
where 2 represents the query and passage memory bank and 4 denotes full precision (4 bytes). 

\begin{table}[h]
\centering
\caption{Comparison of Memory Usage}\label{tab:memory-usage}
    \resizebox{\textwidth}{!}{%
    \setlength{\tabcolsep}{6pt}
        \begin{tabular}{l|c|c|c|c}
        \toprule
        \multirow{2}{*}{\textbf{Method}} & \multirow{2}{*}{\textbf{$N_{\text{local}}/K/N_{\text{total}}/N_{\text{memory}}$}} & \multirow{2}{*}{\textbf{Memory (GB)}} & 
        \multicolumn{2}{c}{\textbf{Additional Memory}} \\
         & & &  Actual & Theoretical \\
        \toprule
        DPR & 8/\phantom{0}1/128/\phantom{000}0 & 7.483 & - & - \\
        \gc & 8/16/128/\phantom{000}0 & 5.158 & - & - \\
        \ga & 8/16/128/\phantom{000}0 & 8.340 & - & - \\
        \modelname & 8/16/128/\phantom{0}128 & 8.342 & 0.002  & 0.0007 \\
        \modelname & 8/16/128/\phantom{0}512 & 8.346 & 0.006  & 0.0029 \\
        \modelname & 8/16/128/1024 & 8.353 & 0.013  & 0.0059 \\
        \modelname & 8/16/128/5096 & 8.382 & 0.042  & 0.0117 \\
        \bottomrule
        \end{tabular}
        }
\end{table}
Moreover, we measured each method's actual memory usage in a VRAM=11GB environment and the results are reported in table \ref{tab:memory-usage}. The results show that \modelname\ uses only up to 0.5\% more memory than \ga, which is a maximum of 12MB even in the largest memory bank size(\(\textbf{N}_\text{memory}=5096\)). This demonstrates that \modelname\ is a memory-efficient method that consumes very limited additional memory compared to \ga. Furthermore, while \gc\ uses less memory than DPR by decomposing complex forward and backward processes, it has the limitation of very slow training speed, as shown in Figure \ref{fig:train speed}.

 \section*{F. License} \label{appendix:license}
The licenses for the assets used in this paper are as follows:
 \begin{itemize}
     \item Overall train code and partial evaluation code from nano-DPR: CC-BY-NC 4.0
     \item Train and evaluation datasets preprocessed by DPR: CC-BY-NC 4.0
     \item Partial evaluation code, and train and evaluation dataset preprocessed by Beir: Apache-2.0
     \item Hard negative score generated by sentence-transformers library: Apache-2.0
 \end{itemize}
\clearpage

\end{document}